\documentclass[twocolumn]{aastex631}
\shorttitle{AASTeX v6.3.1 Sample article}

\usepackage{float}
\usepackage{booktabs}
\usepackage{subfigure}
\usepackage{graphicx}
\usepackage{amsmath}
\usepackage{footnote}
\usepackage{booktabs}
\usepackage{multirow}
\usepackage{threeparttable}
\usepackage{physics}
\usepackage{upgreek}
\usepackage{xspace}
\graphicspath{{images}}

\newcommand{\units}[1]{\ensuremath{\,\mathrm{#1}}}
\newcommand{\erg}{\ensuremath{~\mathrm{erg}}}
\newcommand{\ergs}{\ensuremath{~\mathrm{erg\,s^{-1}}}}
\newcommand{\ergscm}{\ensuremath{~\mathrm{erg\,s^{-1}\,cm^{-2}}}}

\newcommand{\be}{\begin{equation}}
\newcommand{\ee}{\end{equation}}
\newcommand{\ba}{\begin{eqnarray}}
\newcommand{\ea}{\end{eqnarray}}

\newcommand{\mysub}[1]{\ensuremath{_{\mathrm{#1}}}}

\newcommand{\myerror}[2][NONE]{%
  \ifthenelse { \equal {#1} {NONE} } %
  {\ensuremath{\pm #2}}%
  {\ensuremath{_{-#1}^{#2}}}%
}

\newcommand{\cygXt}{Cygnus~X-3\xspace}
\newcommand{\cygXo}{Cygnus~X-1\xspace}
\newcommand{\grs}{GRS~1915+105\xspace}
\newcommand{\sss}{SS~433\xspace}
\newcommand{\vsgr}{V4641~Sgr\xspace}
\newcommand{\maxi}{MAXI~J1820+070\xspace}

\definecolor{dg}{rgb}{0.0, 0.6, 0.1}
\definecolor{ed}{rgb}{1.0, 0.6, 0.1}

\begin{document}

\title{GeV emission around SS 433 with 17 years Fermi-LAT observation}
\author{Qiwang Sun}
\affiliation{School of Physical Science and Technology, Southwest Jiaotong University, Chengdu 610031, People's Republic of China}
\author[0000-0002-7576-7869]{Dmitry Khangulyan}
\affil{Key Laboratory of Particle Astrophysics, Chinese Academy of Sciences, Beijing 100049, People's Republic of China}
\affil{Tianfu Cosmic Ray Research Center, Chengdu 610000, People’s Republic of China}
\email{khangulyan@ihep.ac.cn}
\author[0000-0002-0210-5813]{Jiren Liu}
\affiliation{School of Physical Science and Technology, Southwest Jiaotong University, Chengdu 610031, People's Republic of China}
\author[0000-0003-1039-9521]{Siming Liu}
\affiliation{School of Physical Science and Technology, Southwest Jiaotong University, Chengdu 610031, People's Republic of China}
\email{jrliu@swjtu.edu.cn}

\begin{abstract}
We present an analysis of 17 years of Fermi-LAT observations of the microquasar SS~433. We detect four GeV sources in the region: a newly identified source, PS J1910+0550, located outside W50; the previously reported source J1913+0512; and two features, denoted as the East and West excesses, apparently associated with the X-ray lobes. We focus on the three sources located within W50.

We do not confirm the previously reported periodic modulation from J1913+0512, as no significant periodicity is found in the full 17-year dataset. The East and West excesses exhibit distinct morphological and spectral properties, suggesting different physical origins. The East excess shows a hard spectrum with photon index $\sim1.7$, consistent with inverse Compton emission from relativistic electrons accelerated together with the particles responsible for the X-ray and TeV emission. In contrast, the West excess has a much softer spectrum with photon index $\sim2.6$ and is spatially offset from the known X-ray and TeV emission regions in the western lobe. The spectral shape and offset position of the West excess make it strikingly similar to J1913+0512.

The emission from these two regions can be explained by GeV particles accelerated in SS~433, distributed throughout the source volume, and interacting with localized dense gas targets. Under reasonable assumptions regarding particle transport and energetics, both proton-proton and bremsstrahlung scenarios are viable, although the hadronic scenario is more naturally accommodated. These findings may therefore represent the first observational evidence for the acceleration of cosmic-ray protons in large-scale outflows from Galactic microquasars.
\end{abstract}


\section{Introduction}

Recently LHAASO reported the detection of photons above 20 TeV from five Galactic microquasars \citep[\sss, \grs, \vsgr, \maxi, and \cygXo][]{LH25A}. In four out of these five sources (except \cygXo) the spectrum extends clearly beyond 100~TeV. LHAASO also detected another microquasar, \cygXt,  exclusively in the ultra-high-energy (UHE, $>100$~TeV) regime \citep{LH25B}.
These detections demonstrate that accreting stellar-mass black holes can accelerate particles into the PeV regime \citep[note that the highest detected photon energy from \cygXt reaches $3.7$~PeV][]{LH25B}, and may contribute to Galactic cosmic rays around the `knee' region.
While microquasars were considered as prominent gamma-ray sources \citep[see, e.g.,][]{2009IJMPD..18..347B,2009A&A...497..325B,2011A&A...528A..89B,2017A&A...604A..39D}, these results stimulate significant interests in in particle acceleration and radiation processes in mildly relativistic Galactic jet sources \citep[e.g.][]{Wang25,Zhang25}.

Among the microquasars detected by LHAASO, \sss is a prototypical source, which was first identified as a star with strong H-$\alpha$ emission \citep[from which the commonly used name \sss originates]{1971PW&SO...1a...1S}. Later, it was associated with a variable radio and X-ray source \citep[V1343 Aquilae,][]{1978Natur.276...44C}, and the presence of mildly relativistic outflows was subsequently established \citep{1979ApJ...233L..63M}. The system is formed by an A-type supergiant donor star and a compact object, likely a black hole.  Two precessing jets emerge from the system and extend to distances of about $0.1$~pc from the binary \citep[e.g.][]{2004ApJ...616L.159B}. At larger distances along a direction similar to the precession axis of the inner jets, two X-ray lobes appear on scales of tens of parsecs \citep[e.g.][]{1996A&A...312..306B, Saf22, 2025arXiv251014938S}. The X-ray lobes are embedded within the surrounding radio nebula W50. Recent IXPE observation of the innermost regions of the X-ray eastern lobe of SS 433 confirmed its synchrotron origin and revealed that the magnetic field is preferentially aligned with the jet flow
\citep{2024ApJ...961L..12K}. For recent reviews, see \citet{Fab04,2025arXiv250601106C}.

In the very-high-energy band, HAWC detected two TeV lobes spatially coincident with the X-ray lobes \citep{2018Natur.562...82A,HAWC24}. H.E.S.S. revealed an energy-dependent spatial morphology of SS 433 in 1-30 TeV, with higher-energy photons concentrated closer to the inner regions of the X-ray lobes \citep{HESS24}. They inferred that the very-high-energy emission of SS 433 is produced via inverse Compton scattering by relativistic electrons that lose energy as they propagate outward.
The LHAASO data of SS 433 also show apparent energy-dependent features \citep{LH25A}.
In the 1--25 TeV and 25--100 TeV bands, the emission appears as two point-like sources on either side of SS 433, whereas above 100 TeV the emission is much closer to the center of SS 433. 

Several studies have reported GeV emission around \sss, but the reported results remain controversial.
The analysis of Fermi-LAT observation of \sss was first presented by \cite{2015ApJ...807L...8B}, and subsequently carried out by \cite{2019ApJ...872...25X}, \cite{2019MNRAS.485.2970R}, \cite{2019A&A...626A.113S}. The influence of the nearby bright pulsar J1907+0602 was recognized in later studies and mitigated using a pulsar-gating method by \citet{Fang20} and \citet{Li20}. They found that the brightest feature around \sss was J1913+0515, which is outside the X-ray jets of \sss. The western lobe was detected with a test statistic (TS) of 16, while the eastern lobe with a TS value of about 5 \citep{Fang20}. In addition, \citet{Li20} reported from J1913+0515 a tentative evidence for a periodic modulation with a period of approximately 160 days (which is close to the jet precession period).

\begin{table*}[htbp]
\centering
\setlength{\tabcolsep}{5.5pt}
\renewcommand{\arraystretch}{1.3}

\caption{Spatial-spectral fitting results of excess emission around SS 433 in different energy bands}
\label{table:1}

\begin{tabular}{l c c c c c r }

\toprule
Band & Source & GLON (deg) & GLAT (deg) & Norm ($10^{-12} \text{MeV}^{-1}\text{cm}^{-2}\text{s}^{-1}$) & Index & TS \\
\midrule
\multirow{3}{*}[-1ex]{1--100\,GeV}
& J1913+0512 & $40.07\pm0.03$ & $-2.44\pm0.03$ & $1.45\pm0.30$ & $-2.88\pm0.20$ & 46.08 \\
& East Excess & $39.97\pm0.04$ & $-2.83\pm0.03$ & $0.09\pm0.08$ & $-1.69\pm0.32$ & 15.84 \\
& West Excess & $39.46\pm0.07$ & $-1.88\pm0.07$ & $0.75\pm0.32$ & $-2.58\pm0.31$ & 17.67 \\
\midrule
\multirow{3}{*}[-1ex]{3--100\,GeV}
& J1913+0512 & $40.03\pm0.04$ & $-2.46\pm0.04$ & $3.37\pm2.58$ & $-3.36\pm0.54$ & 21.40 \\
& East Excess & $39.97\pm0.03$ & $-2.84\pm0.04$ & $0.14\pm0.16$ & $-1.81\pm0.41$ & 17.84 \\
& West Excess & $39.44\pm0.05$ & $-1.89\pm0.04$ & $0.93\pm0.69$ & $-2.53\pm0.39$ & 17.97 \\
\bottomrule
\end{tabular}

\vspace{3pt}

\begin{tablenotes}[flushleft]
\item Note: GLON and GLAT are Galactic longitude and latitude, respectively; Norm and index are the power-law model normalization and index; the quoted errors correspond to 68\% confidence levels.
\end{tablenotes}

\end{table*}

Here we investigate the GeV emission around \sss using the full 17-year Fermi-LAT data set. Our goal is to compare the GeV morphology with the recently detected spatially resolved TeV emission and to constrain the underlying particle acceleration and radiation mechanisms.

\begin{figure*}
    \centering
    \includegraphics[trim={0 0.cm 0 0}, clip, width=0.48\textwidth]{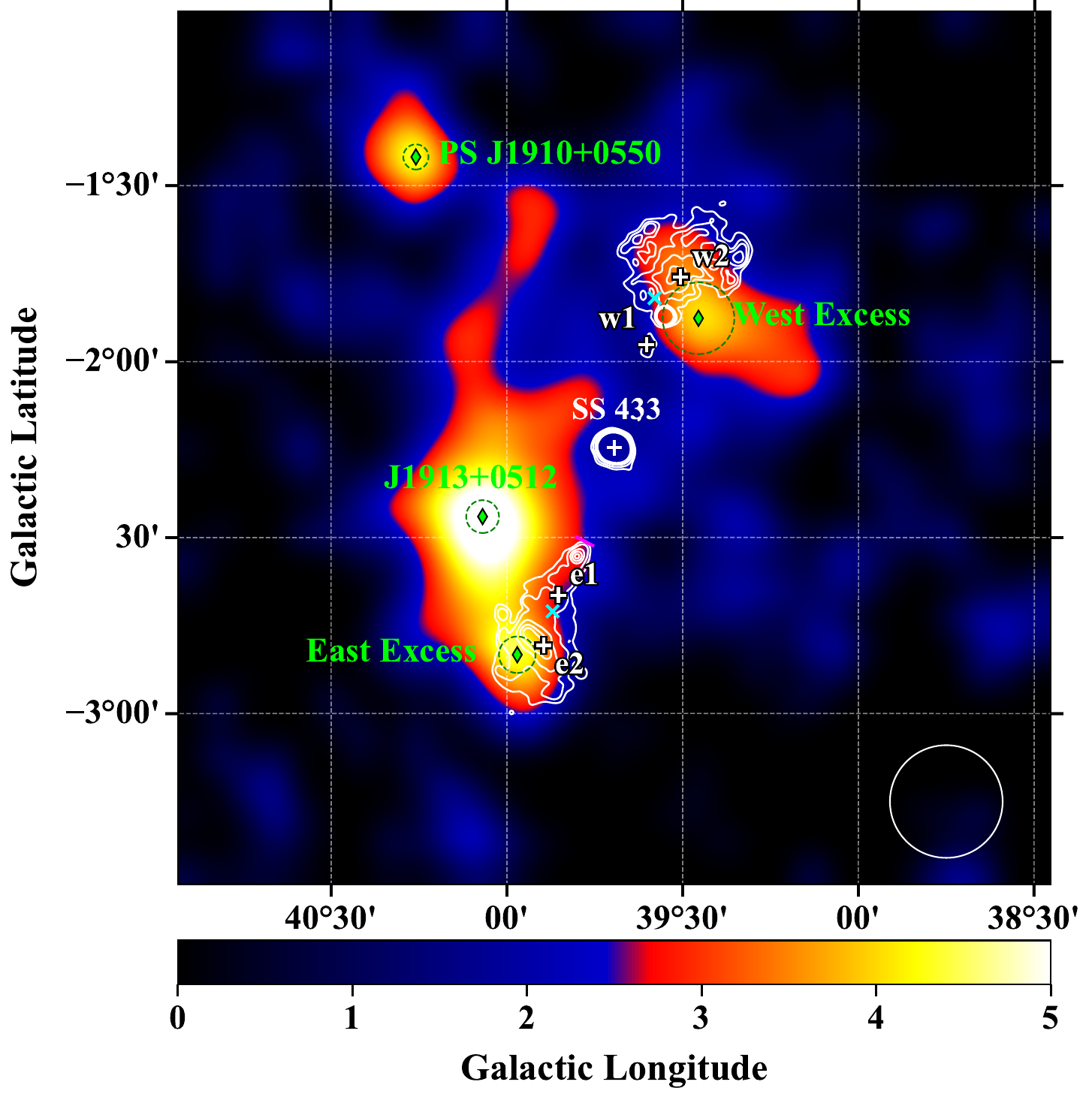}
    \hfill
    \includegraphics[trim={0 0.cm 0 0.0cm}, clip, width=0.48\textwidth]{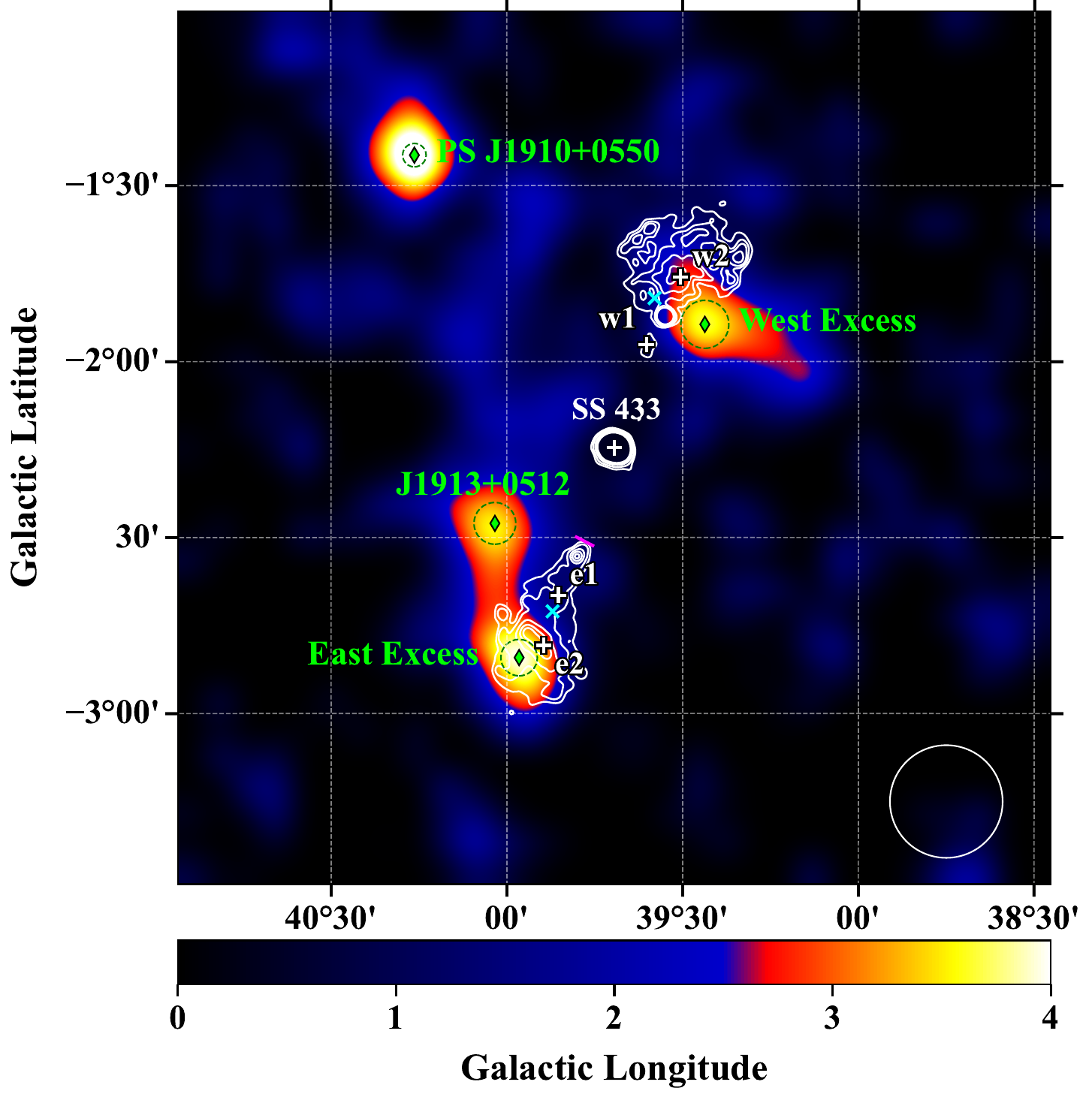}
    \caption{Significance maps ($\sqrt{TS}$) of the emission around SS 433 in the 1–100 GeV (left) and 3–100 GeV (right) bands. The green dashed circles, centered on each source, indicate the 68\% positional uncertainties in the corresponding energy band. The white solid circle in the lower-right corner represents the 68\% containment angle of the Fermi-LAT point-spread function (PSF) at 10 GeV. The magenta short line marks the onset of hard X-ray emission observed by Nustar \citep{Saf22}. The cyan X‑marks indicate the positions where test sources are placed to derive the 95\% upper limits for Fermi‑LAT. The white crosses mark the position of X-ray bright regions in the west ``w1'', ``w2'' and in the east ``e1'', ``e2''.}
    \label{fig:1}
\end{figure*}

\section{Data analysis}

We analyze all the 17 years data around SS 433 observed by Fermi-LAT covering the period up to September 2025.
Since SS 433 is located about 2.2$^\circ$ away from the Galactic disk, we adopt a region of interest (ROI) of 2.5$^\circ\times$2.5$^\circ$ in Galactic coordinates to minimize contamination from the Galactic diffuse emission.
Adopting a larger ROI will not change the main results, but will affect the detection significance a little. 
The spatial resolution of Fermi-LAT is energy-dependent, and is better for higher energy \citep{LAT09}. Therefore, we restrict our analysis to events above 1 GeV and test the impact of different energy selections.
We adopt the standard Galactic diffuse emission model {\tt gll\_iem\_v07.fits} and the isotropic diffuse emission {\tt iso\_p8R3\_SOURCE\_v3\_v1.txt}, together with the fourth source catalog 4FGL-DR4.
The analysis is performed using a binned maximum likelihood fit with a spatial bin size of 0.08 degree and 4 energy bins per decade, with the {\tt P8R3\_Source\_v3} instrument response function. 
Other event selection conditions include a $90^\circ$ zenith cut and a standard quality criteria.

As SS 433 is significantly contaminated by the nearby pulsar PSR J1907+0602, we use the pulsar gating method to suppress its contribution, as did in \citet{Li20}. We adopt the off-pulse phases 0-0.136 and 0.697-1, which correspond to 44\% of the total observation time. The exposure is scaled accordingly. The analysis is performed using Fermitools (version 2.5.1) within FermiPy (version 1.4.1).

\begin{figure}[h]
   \centering
   \includegraphics[trim={0.0cm 0.cm 0 0}, clip, width=0.48\textwidth]{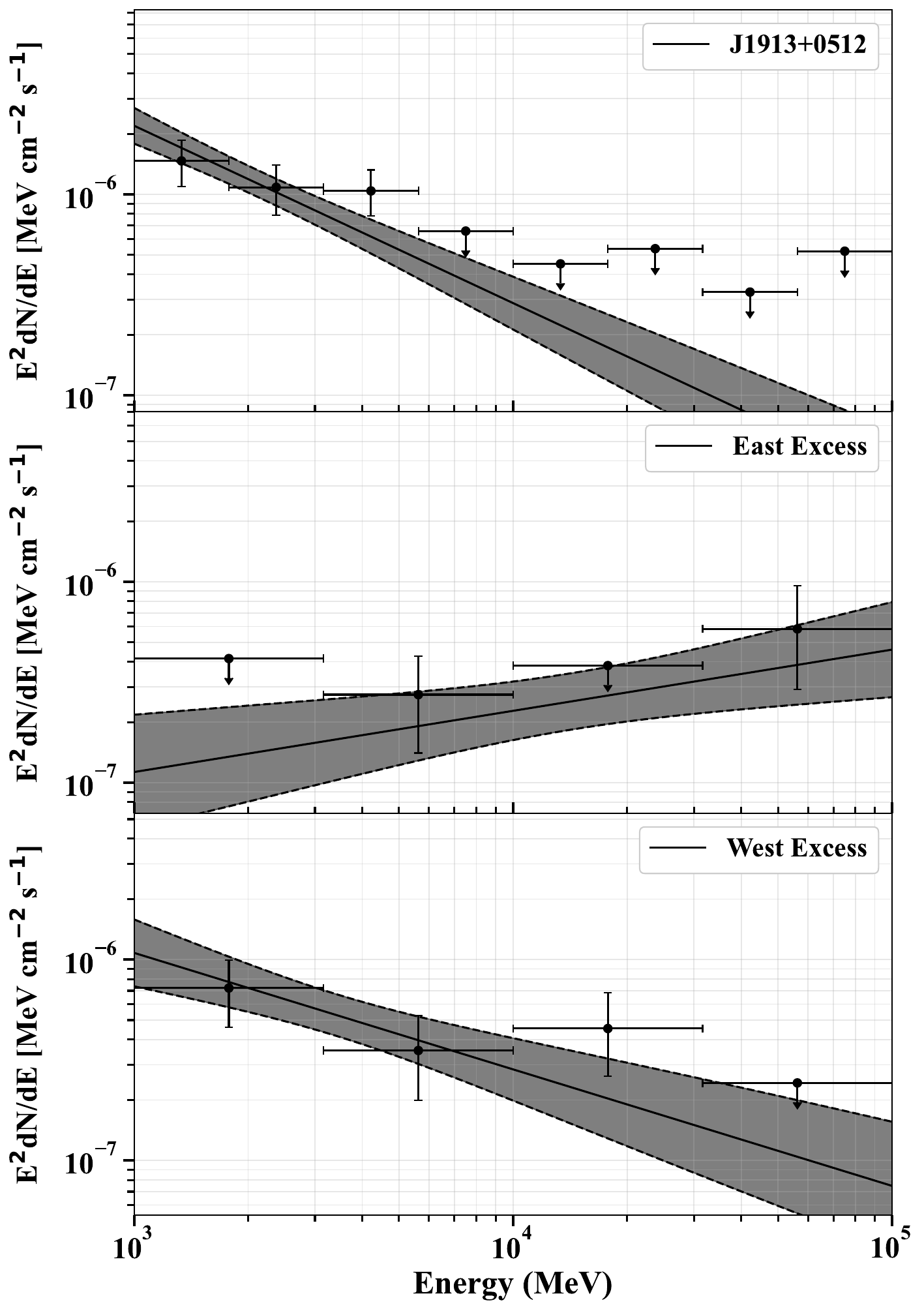}
   \caption{
   Spectral energy distributions (SED) of the excess sources around SS 433. The vertical error bars represent the 68\% confidence intervals of the flux measurements, while the shaded regions around the best-fit curves indicate the 68\% confidence intervals. When the TS value of a flux point is less than 4.0, it is presented as a 95\% confidence upper limit.
   }
   \label{fig:2}
\end{figure}

\section{Analysis results}

\subsection{Morphological study}

The resulting off-pulse TS map around SS 433 in the 1-100 GeV band, assuming an $E^{-2}$ spectrum, is shown in the left panel of Figure 1.  
Several excess features are evident in the vicinity of SS 433.
The most prominent feature is J1913+0512, which is outside the X-ray jet of SS 433 and has been reported in previous works \citep{Fang20, Li20}.
In addition, the excess feature around the western lobe is well separated from J1913+0512 and has also been detected before \citep{Fang20}.
As a surprise, the excess feature around the eastern lobe is also apparent, although it partially overlaps with J1913+0512.
There is also an excess located about 1$^\circ$ north of \sss, labeled as PS J1910+0550 in Figure~\ref{fig:1}. This source is a new source not reported before. It has a significance around 5$\sigma$.
As this source lies outside the radio shell of W50 and is therefore unlikely to be physically associated with SS 433, we do not consider it further in this work.

In the 1-100 GeV band, J1913+0512, the western lobe, and the eastern lobe are detected with TS values of 46, 18, and 16, respectively. These correspond to detection significances of approximately 7$\sigma$ for J1913+0512 and $\sim4\sigma$ for both the western and eastern lobes. The best-fit source positions are summarized in Table 1.

Since the Galactic diffuse emission, as well as the spatial and energy resolution of Fermi-LAT, are all energy-dependent, we 
also tried different energy selection conditions with higher low-energy bounds. The significance map in the 3-100 GeV band is shown in the right panel of Figure 1. As can be seen,
J1913+0512 appears less significant in the 3–100 GeV band compared to the 1–100 GeV band. This is due to the rapid decline of its spectrum at higher energies, as illustrated in Figure 2. In the 3–100 GeV range, the eastern lobe is less affected by contamination from J1913+0512 compared to the 1–100 GeV band. The corresponding TS values in the 3–100 GeV range for J1913+0512, as well as the western and eastern lobes of SS 433, are 21, 18, and 18, respectively. 
The fitted positions are also listed in Table 1. These positions are consistent with those reported in \citet{Fang20} and \citet{Li20}, within uncertainties.

We note that the eastern GeV emission is located outside the bright lenticular X-ray region ``e2", as identified in \citet{Saf97}. No significant GeV excess is detected from either the X-ray ``e1" region or the inner ``head" region of the eastern X-ray jet, as defined by \citet{Saf22} based on Nustar data.
Note that the eastern GeV emission is spatially displaced from the TeV emission associated with the eastern lobe, which is within the ``e2'' and ``e1'' regions as measured by HAWK \citep{HAWC24}, LHAASO \citep{LH25A}, H.E.S.S. \citep{HESS24} and VERITAS \citep{VER26}.

The centroid of western excess shows a noticeable offset from the direction of the western X-ray jet of SS 433. In the 1–100 GeV band, the western emission shows some excess features toward the ``w2" region and the south direction. Notably, the bulk of the western GeV emission lies outside the X-ray contours shown in Figure 1 and is also spatially displaced from the TeV emission, which is aligned well with ``w1" and ``w2" regions. Similarly, J1913+0512 is not spatially coincident with any detected TeV emission.

\subsection{Spectral results}

We fit a power-law model to the energy spectrum of all three sources detected around SS 433 in the 1-100 GeV band.
The spectral fitting results are presented in Figure 2 and summarized in Table 1.
The spectrum of J1913+0512 is quite soft, with a photon index $\sim2.9$. The western lobe of SS 433 exhibits a moderately soft spectrum with an index $\sim2.6$, while the eastern lobe shows a harder spectrum with an index of $\sim1.8$. 
The corresponding fluxes of J1913, the western lobe and the eastern lobe are $2.45 \pm 0.43 \times 10^{-6}$, $1.73 \pm 0.48 \times10^{-6}$, and $1.14 \pm 0.44 \times10^{-6}$ MeV\,cm$^{-2}$\,s$^{-1}$, respectively.

To constrain the GeV emission in regions spatially coincident with the TeV emission, we derive upper limits at the positions reported by \citet{VER26}. The resulting spectral upper limits are presented in Figure 3.

\begin{figure}
    \centering
     \includegraphics[trim={0.0cm 0.0cm 0.0cm 0}, clip, width=0.48\textwidth]
     {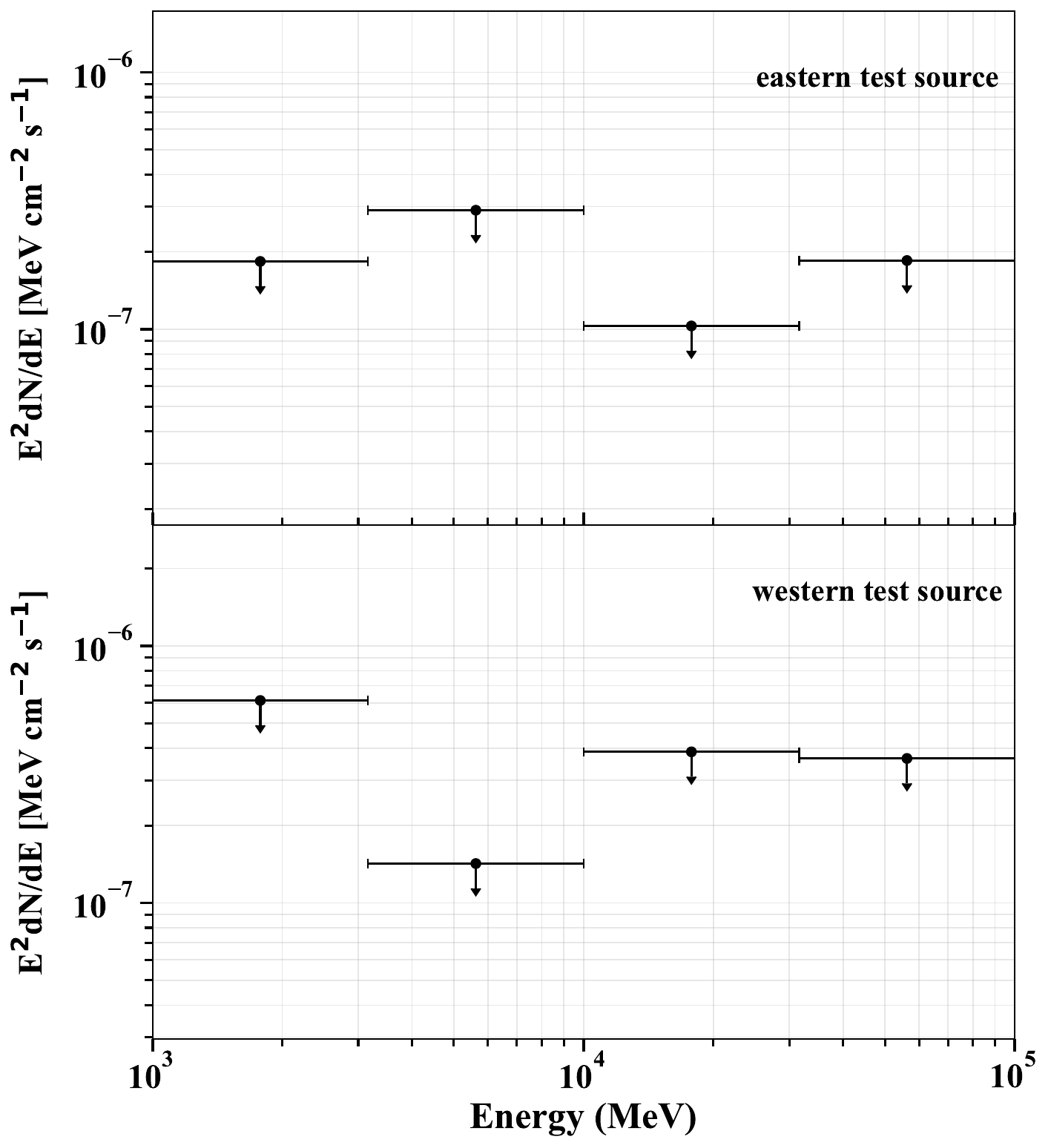}
    \caption{The 95\% spectral upper limits calculated from the eastern (upper panel) and the western (lower panel) test sources. The locations of the test sources are indicated in Figure \ref{fig:1}.}
    \label{fig:3}
\end{figure}
\begin{figure}
    \centering
     \includegraphics[trim={0.0cm 0.0cm 0.0cm 0}, clip, width=0.48\textwidth]
     {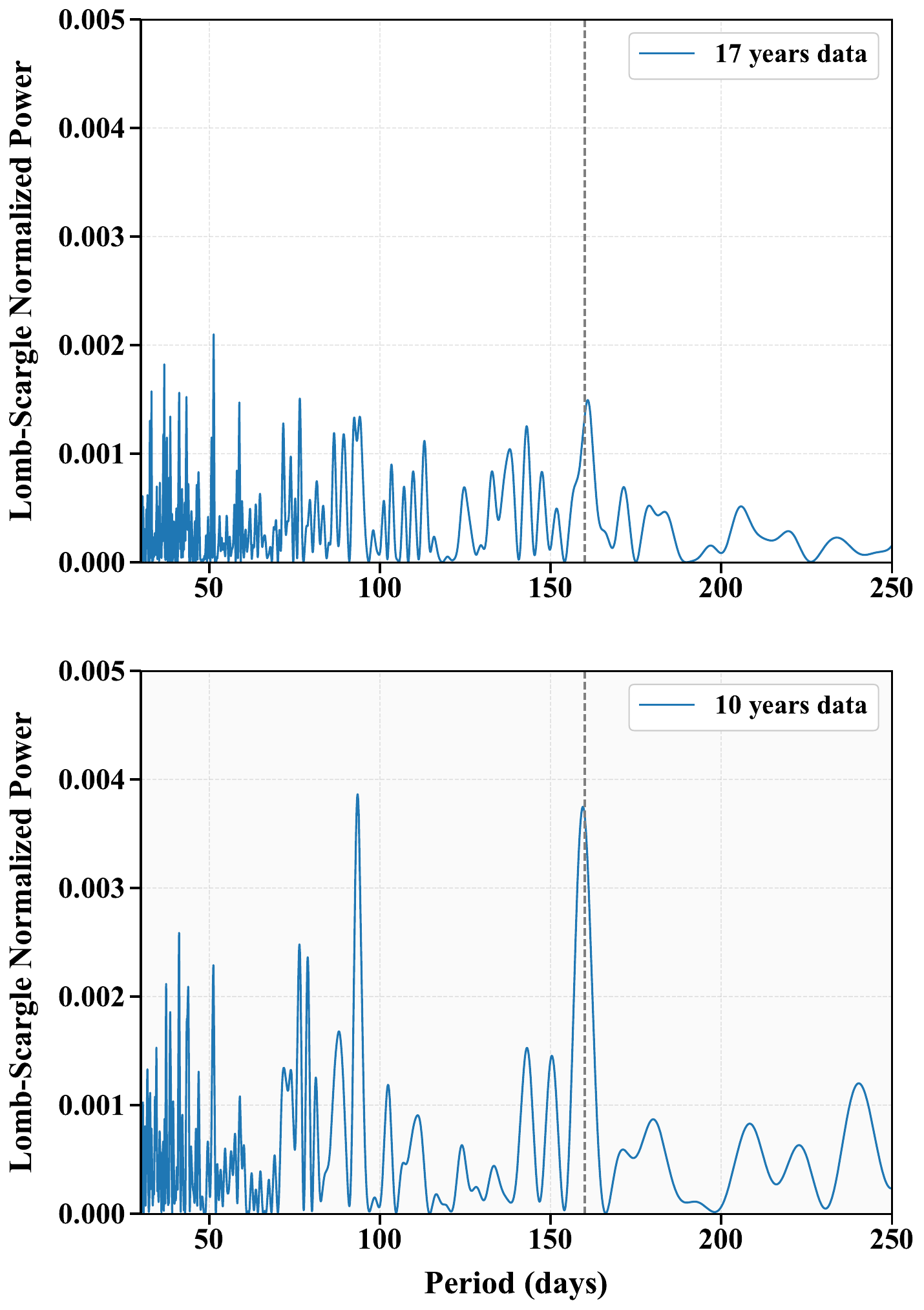}
    \caption{The results of the Lomb–Scargle analysis for the 17-year (upper panel) and 10-year (lower panel) Fermi-LAT data. The vertical gray dashed line marks the 160-day periodic signal reported previously.}
    \label{fig:4}
\end{figure}
\begin{figure}
    \centering
     \includegraphics[trim={0 0.cm 0 0}, clip, width=0.48\textwidth]{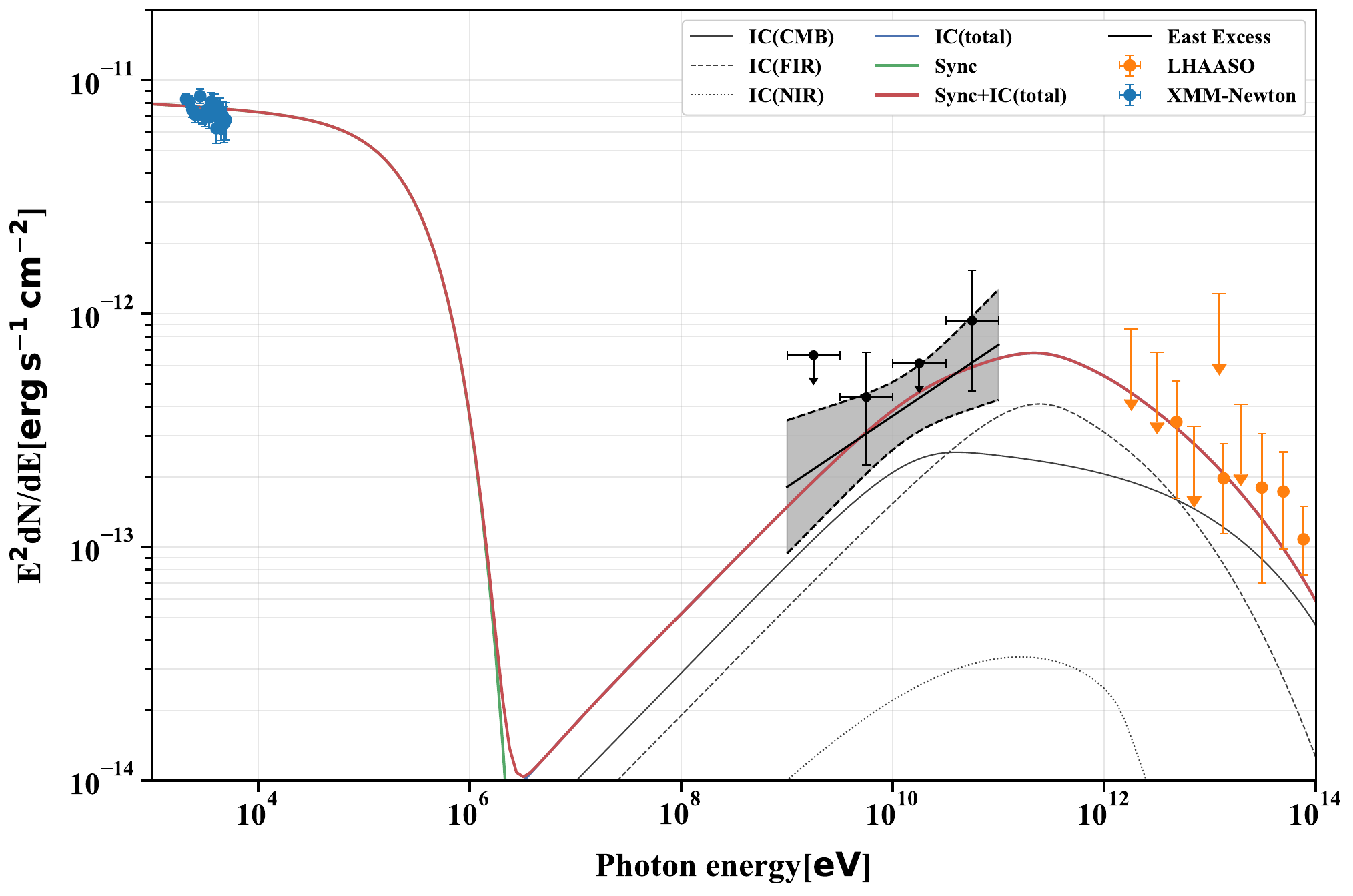}
     \includegraphics[trim={0 0.cm 0 0}, clip, width=0.48\textwidth]{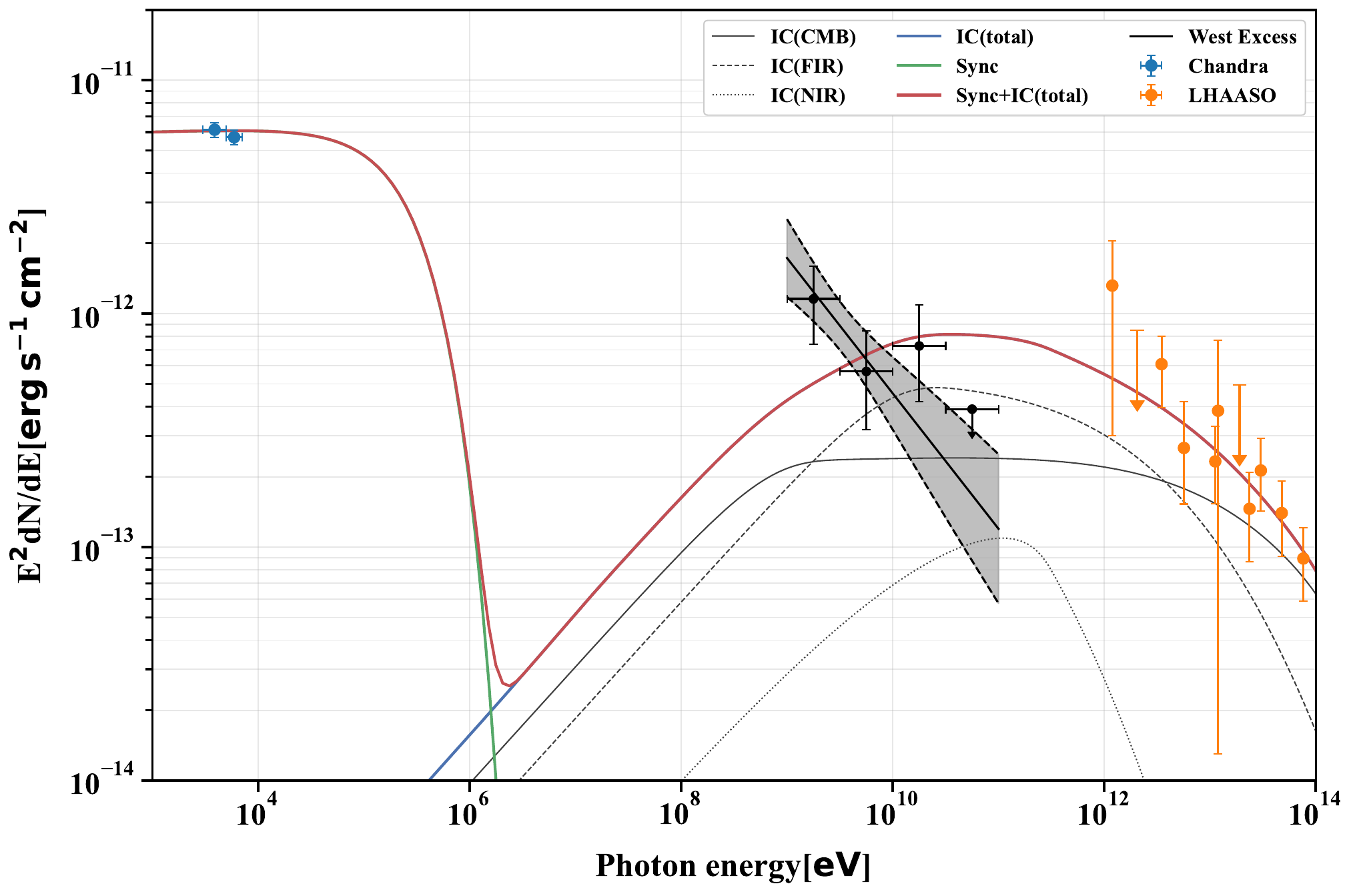}
    \caption{Multiwavelength spectral fitting to the eastern (upper panel) and western (lower panel) lobes of SS 433.}
    \label{fig:placeholder}
\end{figure}

\subsection{Periodicity of Fermi J1913+0512}

\citet{Li20} reported an intriguing periodic signal around 160 days from Fermi J1913+0512 (referred to as J1913+0515 in their work).
To test this result,
we search for periodicity of Fermi J1913+0512 using the 17-year dataset. Following \citet{Li20}, we construct an exposure-corrected, events-weighted light curve of J1913+0512 with 1-day time bins above 1 GeV. The resulting Lomb–Scargle (LS) periodogram is shown in Figure 4. For comparison, we also analyze the same 10-year dataset used by \citet{Li20}, with the corresponding LS result presented in the bottom panel of Figure 4. As shown, the LS power spectrum for the 10-year dataset exhibits a peak around 160 days, consistent with \citet{Li20}. In contrast to the 10-year dataset, the LS power at 160 days in the 17-year dataset is reduced by a factor $\sim2.7$. The corresponding peak is comparable to other fluctuations in the periodogram and no longer stands out as a significant feature. {This suggests that the excess seen in the 10-year dataset is rather a fluctuation than an evidence for real periodicity.}

\subsection{Mutiwavelength Analysis}

{While J1913+0512 appears morphologically separated from the binary system and its non-thermal jets, it is less certain for the emission associated with the eastern and western lobes.  Even if} the observed GeV emission from both the eastern and western lobes {is} not spatially coincident with the dominant X-ray and TeV emission, it is nevertheless instructive to attempt a joint spectral fit. We adopt a simple one-zone leptonic model, similar to that used by \citep{LH25A}, to model the X-ray, GeV, and TeV data for both lobes. This model incorporates electron cooling effects \citep[][Sec. 2.1.4]{LH25A}.
The eastern X-ray data are taken from XMM-Newton \citep{Bri07}, while the western X-ray data are obtained from Chandra \citep{Kay22}. The TeV data are adopted from LHAASO \citep{LH25A}.

The fitting results are presented in Figure 5. The eastern GeV emission, which exhibits a hard spectrum with a photon index below 2, can be reasonably well reproduced by a model with an injection index $\sim2.1$ and a magnetic field strength $\sim19~\mathrm{\upmu G}$.
In contrast, although some data points of the western GeV emission can be described by a similar model, its softer spectrum shape, with a photon index $\sim2.6$, is {impossible} to reproduce with a similar {one-zone} model. This discrepancy suggests that the western GeV emission may arise from a mechanism different from that of the eastern GeV emission.

\section{Discussion and conclusion}

We analyzed the full 17-year Fermi-LAT dataset of the \sss region using pulsar gating to suppress contamination from PSR J1907+0602. We detected four sources: PS J1910+0550, PS J1913+0512, the ``East excess'', and the ``West excess''. Although PS J1910+0550 is a newly discovered source, we focus on the remaining three sources because PS J1910+0550 lies outside the radio nebula W50. The possible association of these three sources with \sss has been discussed in previous studies \citep{Li20,Fang20}. The East and West excesses are located close to the jet lobes of \sss and may represent GeV counterparts of the X-ray and TeV emission regions \citep[see also the joint Fermi-LAT--HAWC analysis in][]{Fang20}. PS J1913+0512 was linked to \sss based on an intriguing periodic modulation reported by \citet{Li20}. Although the significance of the reported periodicity was low, the \(\sim\)160-day period is consistent with the precession period of the inner jets in \sss, lending support to a possible physical connection. The updated dataset provides a $\sim7\sigma$ significance for J1913+0512 and both the eastern and western lobes are detected with $\sim4\sigma$ significance. This enables improved constraints on their positions, morphology, and spectra.

How the precessing jets of \sss can influence the ``heartbeating'' source J1913+0515 (designated as J1913+0512 in this work due to a slight positional shift) remains highly uncertain \citep[see, however,][]{2026ApJ..1001...92O}. We therefore first examined whether the full 17-year dataset strengthens the evidence for the reported periodicity. Figure~\ref{fig:4} shows that, when using a 10-year dataset similar to that analyzed by \citet{Li20}, we recover a low-significance peak at a period of \(\sim\)160 days. However, the updated 17-year analysis reveals no significant periodic modulation in the emission from J1913+0512. This result disfavors the previously reported periodic modulation and weakens the case for an association between J1913+0512 and \sss. Nevertheless, such an association may still be supported by spectral, multi-wavelength, and phenomenological arguments.

The East and West excesses were previously associated with TeV sources through a joint Fermi-LAT--HAWC analysis \citep{Fang20}. However, the full 17-year Fermi-LAT dataset allows us to detect these sources independently, without relying on observations from other instruments, and to obtain robust positional and spectral measurements.

Notably, the East excess lies within the non-thermal lobe but is displaced from the bright lenticular X-ray region ``e2'' \citep{Saf97}, adding a new detail to the gamma-ray morphology of this region. By contrast, both HAWC and LHAASO images of \sss in the 1--25~TeV band place the emission closer to the ``e1'' region. The H.E.S.S. 2.5--10~TeV map of the eastern lobe shows a centroid between the ``e1'' and ``e2'' regions, with only a minor extension beyond ``e2'' \citep{HESS24}, while the 0.8--25~TeV morphology measured by VERITAS is similarly confined within the ``e1'' and ``e2'' regions \citep{VER26}. Taken together, these observations indicate that the detected eastern GeV emission lies outside the bright X-ray emission and is spatially displaced from the bulk of the TeV emission; no significant GeV counterpart is detected from the main TeV-emitting region in the eastern lobe.

The West excess is located about 10$'$ away from the X-ray jet of \sss (at the source distance of \(d=5.5\)~kpc it corresponds to $16$~pc separation), consistent with the position reported by \citet{Li20} and \citet{Fang20}. We also find excess emission extending toward the ``w2'' region and to the south. Since the TeV emission observed by H.E.S.S. and VERITAS is well aligned with the ``w1'' and ``w2'' regions, these results suggest that most of the West excess emission is spatially displaced from both the X-ray and TeV emission sites.

The morphological analysis described above suggests that the East and West excesses may have different physical origins. This interpretation is further supported by their spectral properties. The spectrum of the East excess is hard, with a photon index of $\sim1.7$, whereas the West excess exhibits a much softer spectrum with an index of $\sim2.6$. The energy-dependent gamma-ray morphology favors inverse Compton (IC) scattering as the dominant gamma-ray production channel in \sss \citep{HESS24}, and the hard spectrum of the East excess is consistent with the expected location of the cooling break in the IC spectrum. The confinement time, $T_j$, for relativistic particles in the jet can be estimated as
\be
 T_j \sim \frac{\Delta Z_j}{v_j}\sim 3\times 10^{10}
 \qty(\frac{\Delta Z_j}{30\units{pc}})
 \qty(\frac{v_j}{0.1c})^{-1}\units{s}\,,
\ee
where $\Delta Z_j$ is the length of the jet segment and $v_j$ is the jet speed, which is not necessarily constant along the entire jet \citep[see][for study of jet speed with observations in the X-ray band]{2025ApJ...993L..24T}. Assuming synchrotron losses dominate in a magnetic field of strength $B_j$, the cooling break is expected at
\be
E\mysub{br}\approx 30
\qty(\frac{B_j}{20\units{\upmu G}})^{-2}
\qty(\frac{T_j}{1\units{kyr}})^{-1}\units{TeV}\,.
\ee
Estimating the corresponding cooling break in the IC spectrum requires adopting a target photon field. Since, for comparable energy densities, lower-temperature photon fields can provide the dominant IC contribution for hard electron spectra \citep{1997MNRAS.291..162A}, we consider the cosmic microwave background (CMB) with $T_{\rm CMB}=2.7\,\mathrm{K}$. Using the relations from \citet{2014ApJ...783..100K}, the energies of gamma-ray photons can be related approximately to those of their parent electrons. For the CMB target field, $30$~TeV electrons  predominantly produce gamma-ray emission at $2$~TeV. This estimate indicates that radiative cooling is negligible for electrons responsible for the GeV emission, whereas the TeV emission should be significantly affected by cooling.

This simple estimate has different implications for the East and West excesses. In the case of the East lobe, all three observed features of the GeV emission, namely (i) the hard spectrum, (ii) the absence of bright emission near the likely particle acceleration sites, and (iii) the accumulation of emission beyond the regions bright in the TeV band, are consistent with an IC origin of the gamma-ray emission (see Figure~\ref{fig:placeholder}, top panel). In contrast, the spectral properties of the West excess strongly disfavor an IC origin, at least within the framework of a scenario that allows simultaneous reproduction of the TeV and X-ray data (see Figure~\ref{fig:placeholder}, bottom panel).

Two other radiation mechanisms can potentially account for the West excess: proton-proton ($pp$) interactions and electron bremsstrahlung ($bs$). In both cases, the target field is provided by the ambient gas with density $n_{\mathrm{gas}}$, and the corresponding cooling times are of the same order,
\be
t_{\mathrm{pp/bs}} \sim 10^{15}\qty(\frac{n_{\mathrm{gas}}}{1\units{cm^{-3}}})^{-1}\units{s}\,.
\ee
Therefore, the dominant emission channel, $pp$ or $bs$, is determined primarily by whether relativistic protons or electrons carry the larger fraction of non-thermal energy. The detection of line emission from the base of the jet indicates the presence of ions in the mildly relativistic outflow. Therefore, if particle acceleration occurs in the presumed acceleration sites located in the innermost regions of the X-ray lobes, both electrons and protons are expected to be accelerated.

The jet density can be estimated as
\be
\begin{split}
  n_{\mathrm{gas}} &\lesssim \frac{2L_k}{\pi R_j^2 m_p v_j^3}\\
        &\approx 6\times10^{-5}
        \qty(\frac{L_k}{10^{39}\ergs})
        \qty(\frac{R_j}{5\units{pc}})^{-2}
        \qty(\frac{v_j}{0.1c})^{-3}
        \units{cm^{-3}}\,.
\end{split}
\ee
Here $m_p$ is the proton mass, $c$ is the speed of light, and $L_k$ and $R_j$ denote the jet kinetic luminosity and jet radius, respectively. This low density effectively rules out production of the detected GeV emission within the jet through either the $pp$ or $bs$ channels.

Thus, the detected West excess, and possibly also J1913+0512 if its association with \sss is confirmed, point to the presence of relativistic particles outside the TeV/X-ray lobes. These particles may, for example, be accelerated together with the particles responsible for the formation of the lobes, but owing to their much longer cooling timescales they can escape from the lobes into the surrounding cocoon and farther away from the system. The confinement time of GeV particles can be very long, potentially exceeding the source age, $T_{\mathrm{age}}\sim30$~kyr. Since radiative cooling is also negligible, the total energy stored in non-thermal particles can be estimated as
\be
\begin{split}
  E_{\mathrm{nt}} &= \kappa_{\mathrm{nt}} \bar{L}_k T_{\mathrm{age}}\\
                  &\approx10^{49}
                  \qty(\frac{\kappa_{\mathrm{nt}}}{10^{-2}})
                  \qty(\frac{\bar{L}_k}{10^{39}\ergs})
                  \qty(\frac{T_{\mathrm{age}}}{30\units{kyr}})
                  \erg\,,
\end{split}
\ee
where $\kappa_{\mathrm{nt}}$ is the fraction of jet power transferred to the relevant non-thermal particles, and $\bar{L}_k$ is the time-averaged jet power.

Since diffusion is expected to produce a fairly homogeneous particle distribution, the detection of localized sources requires an enhanced target density in those regions. This disfavors an IC origin for the West excess and J1913+0512, and instead points to $pp$ and $bs$ processes in the hadronic and leptonic scenarios, respectively. The corresponding gamma-ray flux can be estimated as
\be\label{eq:flux}
\begin{split}
  F_{\mathrm{GeV}} &= \frac{f E_{\mathrm{nt}}}{4\pi d^2 t_{pp/bs}}\\
                   &\approx8\times10^{-13}\ergscm\times \\
                   &\qty(\frac{f}{10^{-3}})
                   \qty(\frac{E_{\mathrm{nt}}}{10^{49}\erg})
                   \qty(\frac{n_{\mathrm{gas}}}{3\times10^{2}\units{cm^{-3}}})\,.
\end{split}
\ee
Here $f$ is the fraction of non-thermal particles contributing to the source (i.e., the filling factor).

The flux levels of the West excess and J1913+0512 are approximately $10^{-12}\ergscm$, comparable to the estimate provided by Eq.~\eqref{eq:flux}. Let us therefore examine whether the adopted parameter values are physically plausible. First, modeling of the X-ray and TeV emission from the lobes suggests that only a relatively small fraction of the jet power is transferred to relativistic electrons, $\kappa_{\mathrm{nt}}\approx 10^{-3}$. Thus, if the particles responsible for the emission outside the lobes were accelerated at the same sites, the expected flux in the leptonic scenario would be lower by approximately one order of magnitude. In contrast, a significantly larger fraction of energy may be transferred to relativistic protons, $\kappa_{\mathrm{nt}}\approx 10^{-1}$, substantially relaxing the requirements of the $pp$ scenario.

To estimate the filling factor $f$, one needs to determine the spatial distribution of GeV particles within the source. This distribution can be affected by different transport processes as well as by the full evolutionary history of the system, owing to the long cooling times of GeV particles. The structure observed in the radio and X-ray bands is shaped by hydrodynamic processes and can be approximated as an ellipsoid with characteristic radii of $50$--$100\units{pc}$. In highly perturbed plasma, particle transport may proceed in the Bohm diffusion regime, with a diffusion coefficient
\be
\begin{split}
  D_{b}&= \frac{cE}{3eB}\\
       &\approx 2\times10^{22}
       \qty(\frac{E}{10\units{GeV}})
       \qty(\frac{B}{20\units{\upmu G}})^{-1}
       \units{cm^2\,s^{-1}}\,.
\end{split}
\ee
The corresponding Bohm diffusion length over the age of the source, $l_{bd}=\sqrt{6D_bT_{\mathrm{age}}}$, is very small, $l_{bd}\ll1$~pc. If particle transport is significantly faster, for example with a diffusion coefficient comparable to the typical Galactic value, then the effective diffusion coefficient would be approximately $10^6$ times larger, allowing multi-GeV particles to occupy a region with a radius of order $100$~pc. Thus, while Bohm diffusion predicts propagation scales much smaller than the source size, faster diffusion may allow particles to propagate throughout the entire source. Consequently, the spatial distribution of non-thermal particles remains highly uncertain.

The projected positions of the West excess and J1913+0512 both lie within the source structure, at angular separations of $\sim10'$ and $\sim20'$ from ``w1'' and ``e1'', respectively. These offsets correspond to projected distances of $16\units{pc}$ and $32\units{pc}$, respectively. This suggests that GeV particles may be distributed throughout the entire source volume, and the filling factor can therefore be roughly estimated as $(3.8\units{pc}/50\units{pc})^3\approx 4\times10^{-4}$, where we adopt the molecular cloud size inferred at the position of J1913+0512 \citep{2026ApJ..1001...92O}.

Although the gas density affects both channels in a similar way, it is important to adopt a reasonable fiducial value, as this may help discriminate between the two scenarios. \citet{2026ApJ..1001...92O} studied the vicinity of J1913+0512 and suggested the presence of a molecular cloud with a mean density of $10^{2.5}\units{cm^{-3}}$ and a radius of $R_c\approx 3.8\units{pc}$. Although no estimates of the gas density are currently available for the West excess region, we cannot exclude a similar density there.

Summarizing these arguments, we conclude that the West excess and J1913+0512 can be explained by GeV particles distributed throughout the source under reasonable assumptions regarding their acceleration and transport. The $pp$ scenario may nevertheless be more favorable because a larger fraction of the jet power can be transferred to relativistic protons than to electrons during acceleration in mildly relativistic outflows. Under this assumption, the properties of the molecular cloud at the position of J1913+0512 readily explain the observed flux level. By contrast, realization of the leptonic $bs$ scenario may require more efficient acceleration of low-energy electrons than inferred from modeling of the X-ray and TeV emission and/or a highly non-homogeneous distribution of GeV electrons within the source.

\section*{Acknowledgements}
We thank Matthew Kerr for kindly providing us the updated ephemeris of PSR J1907+0602 and Haotian Wang for helpful discussions.
We acknowledge the support by the National Natural Science Foundation of China (12473044) and by the fundamental research funds for the central universities (2682025CX065). DK acknowledges support by RSF grant No. 24-12-0045.

\end{document}